\documentclass[multphys,vecphys]{svmult}

\usepackage{makeidx}         
\usepackage{graphicx}        
\usepackage{multicol}        
\usepackage[bottom]{footmisc}

\makeindex             

\begin{document}

\title*{SINFONI's take on Star Formation, Molecular Gas, and Black
  Hole Masses in AGN}
\titlerunning{Star Formation, Molecular Gas, \& Black Holes Masses in AGN}

\author{R.~Davies, R.~Genzel, L.~Tacconi, F.~M\"uller~Sanchez, 
J.~Thomas, \and S.~Friedrich}
\authorrunning{R. Davies et al.} 

\institute{Max Planck Institut f\"ur extraterrestrische Physik,
  Postfach 1312, 85741, Garching, Germany}
%
%
\maketitle

\begin{abstract}
We present some preliminary (half-way) results on our adaptive optics
spectroscopic survey of AGN at spatial scales down to 0.085$''$.
Most of the data were obtained with SINFONI which provides integral
field capability at a spectral resolution of $R\sim4000$.
The themes on which we focus in this contribution are: star formation
around the AGN, the 
properties of the molecular gas and its relation to the torus, and the
mass of the black hole.
\end{abstract}

\section{The AGN Sample}
\label{dav:sec:sample}

The primary criteria for selecting AGN were that 
(1) the nucleus should be bright enough for adaptive optics
correction, 
(2) the galaxy should be close enough that small spatial scales can be
resolved, and 
(3) the galaxies should be ``well known'' so that complementary
data can be found in the literature.
These criteria were not applied strictly, since some targets were also
of particular interest for other reasons.
The resulting sample of 9 AGN is listed in Table~\ref{dav:tab:sample}.
The observations of these are now completed, and while the data for
some objects has been fully analysed, others are still in a
preliminary stage.
Additional AGN will likely be added once the Laser Guide Star Facility
is commissioned.

\begin{table}
\begin{centering}
\caption{AGN sample}
\label{dav:tab:sample}       
\begin{tabular}{llrlrll}
\hline\noalign{\smallskip}
Target & Classification & Dist. & \ \ \ & \multicolumn{3}{c}{Observations} \\
       &                & (Mpc)    && Date & \ \ \ & Instrument\\
\noalign{\smallskip}\hline\noalign{\smallskip}
Mkn 231$^1$  & ULIRG / Sy 1 / QSO & 170 && May '02 && Keck / NIRC2 \\
NGC 7469$^2$ & Sy 1 & 66 && Nov '02 && Keck / NIRSPAO \\
IRAS 05189-2524 \ \ & ULIRG / Sy 1 & 170 && Dec '02 && VLT / NACO \\
Circinus$^3$ & Sy 2 & 4 && Jul '04 && VLT / SINFONI \\
NGC 3227$^4$ & Sy 1 & 17 && Dec '04 && VLT / SINFONI \\
NGC 3783 & Sy 1 & 42 && Mar '05 && VLT / SINFONI \\
NGC 2992 & Sy 1 & 33 && Mar '05 && VLT / SINFONI \\
NGC 1068 & Sy 2 & 14 && Oct '05 && VLT / SINFONI \\
NGC 1097 & LINER / Sy 1 & 18 && Oct '05 && VLT / SINFONI \\
\noalign{\smallskip}\hline
\end{tabular}
\end{centering}
$^1$ Davies et al. 2004a \cite{dav:dav04a};
$^2$ Davies et al. 2004b \cite{dav:dav04b};
$^3$ M\"uller Sanchez et al. 2006 \cite{dav:mul06};
$^4$ Davies et al. 2006 \cite{dav:dav06};
\end{table}

\begin{figure}
\centering
\includegraphics[height=4.5cm]{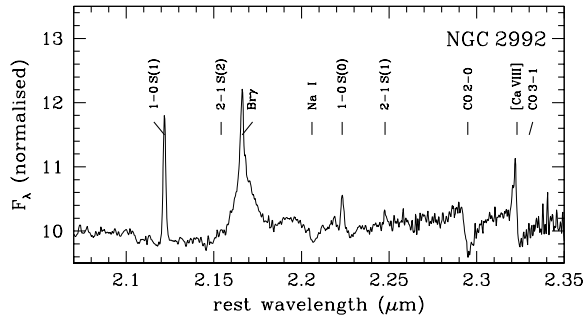}
\caption{Spectrum of the central 0.5$''$ of NGC\,2992, taken with
  SINFONI at a spatial resolution of 0.3$''$ and a spectral
  resolution of $R\sim3400$.
The stellar absorption features are clear, as is the coronal 
[Ca\,{\sc VIII}], the H$_2$ 1-0\,S(1), and both narrow and broad
  Br$\gamma$.}
\label{dav:fig:n2992}       
\end{figure}

One immediate result, which has a bearing on the classifications in
the table, is the frequent detection of broad Br$\gamma$ --
i.e. with FWHM at least 1000\,km\,s$^{-1}$.
An example of this is given in Fig.~\ref{dav:fig:n2992}.
In only 3 galaxies was no broad Br$\gamma$ detected: Circinus,
NGC\,1068, and NGC\,1097 (in which even the narrow Br$\gamma$ is so
weak that it is almost lost in the stellar absorption features).

\section{Star Formation}
\label{dav:sec:starform}

The topics we address here are the spatial scales on which
stars exist around the AGN, the age and star formation history of
these stars, and their contribution to the bolometric luminosity with
respect to that of the AGN itself.

\begin{figure}
\centering
\includegraphics[height=11cm]{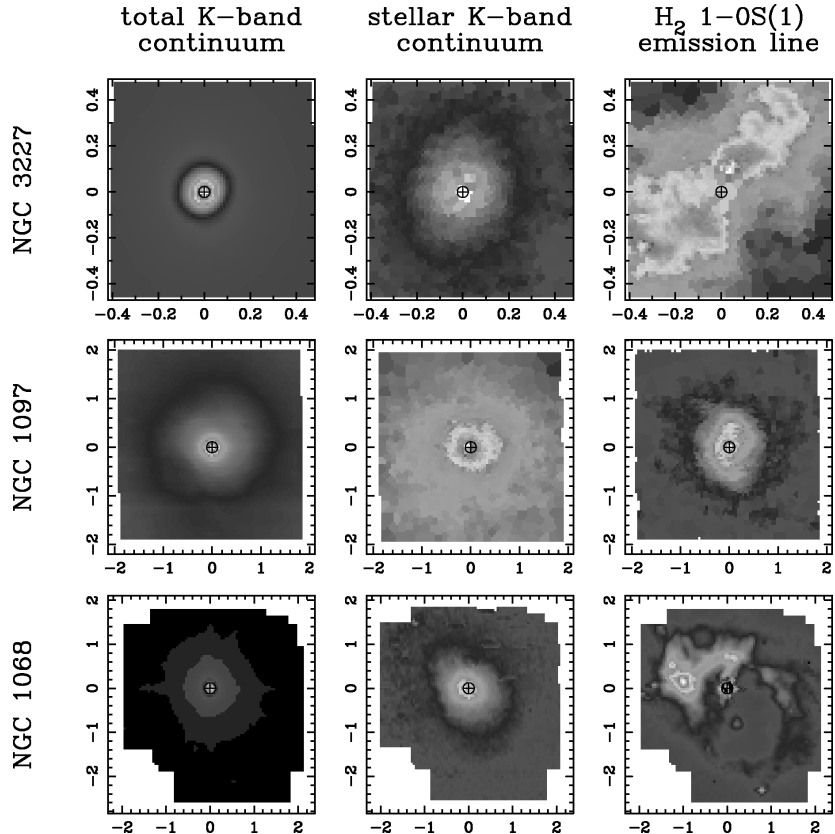}
\caption{Images from SINFONI of the 3 AGN NGC\,3227, NGC\,1097, and
  NGC\,1068, which are at approximately the same
  distance so that 1$''$ $\sim$ 70--80\,pc.
The left panels show the full continuum at 2.1$\mu$m; 
the centre panels the stellar K-band continuum (derived from the CO
  bandheads); and the right panels the H$_2$ 1-0\,S(1) line emission.
}
\label{dav:fig:plot3}       
\end{figure}

The stellar K-band (or equivalently H-band) continuum can be
distinguished from the non-stellar continuum
via the depth of stellar absorption features such as the CO bandheads,
because for any ensemble of stars the intrinsic depth will not vary
much once late-type stars appear (see Davies et al. 2006 \cite{dav:dav06}
for a more detailed discussion of this).
Doing so immediately allows one to assess the physical size
scale of the stellar population close to the AGN (see Fig.~\ref{dav:fig:plot3}).
In addition it permits a lower limit to be put on 
the bolometric luminosity originating in stars.
This is because, while a stellar population which is still
forming stars will have $L_{\rm bol}/L_{\rm K} \sim 50$ (or even higher
if it is very young), even an old passively evolving population has 
$L_{\rm bol}/L_{\rm K} \sim 20$.
In most cases we are able to apply tighter constraints than this by
considering other diagnostics.
For example, from the morphology and kinematics one can estimate the
fractions of the narrow Br$\gamma$ flux that are associated with stars
and with the AGN's narrow line region.
Similarly, it is often possible to estimate the fractions of the
radio continuum associated with the AGN and stars:
the former will be unresolved and have very high brightness
temperatures (see Condon et al. 1991 \cite{dav:con91}).
The ratio of either of these to the stellar K-band continuum can provide
strong constraints on the star formation time scales and hence the
bolometric luminosity from stars close around the AGN.

Our preliminary results are:
\begin{itemize}

\item
In all 9 cases we have resolved a stellar population around the AGN on
the scales we have achieved (0.08--0.3$''$); and the stellar
luminosity increases as one approaches the AGN.

\item
In the 5 cases we have analysed in detail so far (Mkn\,231, NGC\,7469,
IRAS\,05189-2524, Circinus, NGC\,3227), the stellar
population is young: the range of ages we find is 40--120\,Myr 

\item
The (young) stellar luminosity is comparable to that of the AGN on
scales of 1\,kpc (Mkn\,231, IRAS\,05189-2524); 
is 10--50\% of the AGN on scales of 50--100\,pc (NGC\,7469, NGC\,3227); 
and is a few percent of the AGN on scales of 10--20\,pc (Circinus). 

\end{itemize}

\section{Molecular Gas}
\label{dav:sec:torus}

The H$_2$ morphologies traced by the 1-0\,S(1)
line show a much greater diversity than the stellar distributions, as
typified in Fig.~\ref{dav:fig:plot3}.
This might be expected since it is known that distribution of gas is
strongly influenced by dynamical resonances and outflows.
However, when analysing the morphologies on $\sim$10\,pc scales, one
needs to remember that the 1-0\,S(1) line traces only hot
(typically 1000--2000\,K) gas, and hence the very local environment
will have an important impact on the observed luminosity
distribution: for example, is there
a particularly massive star cluster nearby or has there been a recent
supernova?
With this caveat in mind, our preliminary results are:
\begin{itemize}

\item
the 1-0\,S(1) emission is stronger closer to the AGN (with
the exception of NGC\,1068) indicating the gas distribution is also
concentrated towards the nucleus on scales of 10--50\,pc.

\item
the kinematics show ordered rotation (again excepting NGC\,1068) but
also remarkably high velocity dispersion -- in the
range $\sigma = 70$--140\,km\,s$^{-1}$, giving 
$V_{\rm rot}/\sigma \sim 1$. 
This means that the gas must be rather turbulent, most likely due to
heating from the AGN and/or star formation, and as a result is
probably geometrically thick.

\item
Given the size scales on which models predict the molecular torus
around AGN should exist (10--100\,pc, e.g. most recently Schartmann et
al. 2005 \cite{dav:sch05}), and the fact that the torus
must have a large enough scale height to collimate ionisation cones,
it is reasonable to propose that the gas we have seen in these data is
associated with the torus.

\end{itemize}

\section{Black Hole Masses}
\label{dav:sec:bhmass}

Since it was first discovered, the
relation between the mass of the supermassive 
black hole $M_{\rm BH}$ and the velocity
dispersion $\sigma_*$ of the surrounding spheroid has become a
cornerstone of galaxy evolution and black hole growth in the
cosmological context.
However, almost without exception the `reliable' black hole masses
(typically based on stellar 
kinematics and resolving the black hole's radius of influence) have
been derived only for nearby bulge 
dominated E/S0 quiescent galaxies (see the review by Ferrarese
\& Ford 2005 \cite{dav:fer05}).
While extremely challenging, it is therefore crucial to determine
stellar dynamical black hole masses in AGN -- not only to verify
that the $M_{\rm BH} - \sigma_*$ relation holds for galaxies which
are by definition active, but to assess its scatter for these
galaxies, and to provide a comparison to reverberation masses which
might then allow one to constrain the geometry of the broad line region.

The high spatial resolution and integral field capability of SINFONI
provide an ideal combination to do this, and we have successfully derived 
$M_{\rm BH}$ in NGC\,3227 from stellar kinematics -- the first time
for a Seyfert~1 -- using Schwarzschild orbit superposition techniques. 
Details of the specific code, which is based on that used by the
Nuker team, are given in Thomas et al. (2004) \cite{dav:tho04}.
While the inclination and mass-to-light ratios are often uncertain
parameters, for NGC\,3227 they are relatively well constrained.
Nevertheless, we have explored the range of values which the modelling
would permit and find it to be consistent with those expected,
giving us confidence that the results are physically meaningful and
reasonably robust.
The resulting range of permissible black hole masses is 
$M_{\rm BH} = 5\times10^6$--$2\times10^7$\,$M_\odot$.

The range is a result of the degeneracy between the black hole mass
and the `effective' mass-to-light ratio of the stellar population,
which includes the contribution of the gas mass.
If the gas is significantly less concentrated than the stars, then the
higher $M_{\rm BH}$ is possible; 
on the other hand if the gas is strongly centrally concentrated in a
similar way to the stars, then $M_{\rm BH}$ must be correspondingly
lower.

That the mass we find is within a factor of 2--3 of the masses found
by other methods suggests that all are satisfactory to this level of
accuracy.
However, the fact that the mass is also likely to be a factor of a few
below that implied by the $M_{\rm BH} - \sigma_*$ relation, while in
contrast the stellar dynamical mass of Cen\,A (Silge et al. 2005,
\cite{dav:sil05}) is a factor of several greater, may indicate that for AGN
the scatter around this relation could be very considerable.

%


\printindex
\end{document}